# MORPHOMETRIC ANALYSES OF THE VISUAL PATHWAYS IN MACULAR DEGENERATION

Aditya T. Hernowo[a,b] [1], Doety Prins[a], Heidi A. Baseler[c, e], Tina Plank[d],
Andre Gouws[5], Johanna M.M. Hooymans[f], Antony B. Morland[c, e],
Mark W. Greenlee[d] and Frans W. Cornelissen[a]

## ABSTRACT

**Introduction**. Macular degeneration (MD) causes central visual field loss. When field defects occur in both eyes and overlap, parts of the visual pathways are no longer stimulated. Previous reports have shown that this affects the grey matter of the primary visual cortex, but possible effects on the preceding visual pathway structures have not been fully established.

**Method**. In this multicentre study, we used high-resolution anatomical magnetic resonance imaging and voxel-based morphometry to investigate the visual pathway structures up to the primary visual cortex of patients with age-related macular degeneration (AMD) and juvenile macular degeneration (JMD).

**Results**. Compared to age-matched healthy controls, in patients with JMD we found volumetric reductions in the optic nerves, the chiasm, the lateral geniculate bodies, the optic radiations and the visual cortex. In patients with AMD we found volumetric reductions in the lateral geniculate bodies, the optic radiations and the visual cortex. An unexpected finding was that AMD, but not JMD, was associated with a reduction in frontal white matter volume.

**Conclusion**. MD is associated with degeneration of structures along the visual pathways. A reduction in frontal white matter volume only present in the AMD patients may constitute a neural correlate of previously reported association between AMD and mild cognitive impairment.

[1] *The first and second author contributed equally to this work*
[a] *Laboratory for Experimental Ophthalmology, University Medical Center Groningen, Groningen, the Netherlands*
[b] *Department of Basic Medical Sciences, University Malaysia Sarawak, Kuching, Malaysia*
[c] *Department of Psychology, University of York, York, United Kingdom*
[d] *Institute for Experimental Psychology, University of Regensburg, Regensburg, Germany*
[e] *Centre for Neuroscience, Hull-York Medical School, York, United Kingdom*
[f] *Department of Ophthalmology, University Medical Center Groningen, Groningen, the Netherlands*

*The first author was supported by the "RuG Fellowship Program" grant scheme from the University of Groningen, the Netherlands. Graduate School of Medical Sciences (GSMS) supported the second author. This work was further supported by research grants from Stichting Nederlands Oogheelkundig Onderzoek (SNOO), Nelly Reef Fund, and – via UitZicht – Stichting MD Fonds, Landelijke Stichting voor Blinden en Slechtzienden (LSBS), and Algemene Nederlandse Vereniging ter Voorkoming van Blindheid (ANVVB) to FWC, the German Research Foundation (DFG: Project FOR 1075, TP8) to TP and MWG, and the UK Medical Research Council to ABM. The funding organizations had no role in the design or conduct of this research.*

# INTRODUCTION

Macular degeneration (MD) is a class of eye diseases that causes visual field defects that are located in or near the central visual field (central scotoma). According to the World Health Organization, in 2002, age-related macular degeneration (AMD) and the juvenile-type macular degeneration (JMD) were the causes of blindness and low-vision in 8.7% of the 160 million cases worldwide.[1] In recent reports, AMD prevalence ranged from approximately 3 - 12%[2-9], while the prevalence of JMD was 0.03%.[10] In Europe, the retinal pathologies underlying AMD and JMD are listed in the top five of causes of visual loss in both children and adults.[11] In the European adult population, AMD is the most common cause of visual impairment.[12]

Visual field loss in both AMD and JMD is primarily due to a loss of photoreceptors. Despite this, and the similarity in the resulting visual field defects, AMD and JMD do not share a common disease mechanism. The various macular pathologies underlying JMD may start early in childhood and are mostly inherited. They include diseases such as Stargardt's disease, Best's vitelliform retinal dystrophy (Best's disease), cone-rod dystrophy, and central areolar choroidal dystrophy. In AMD, the accumulation of sub-macular deposit called drusen compromises retinal metabolism and leads to the degeneration of the macula with or without neovascularization.[13-16]

An absence of afferent or efferent stimulation may result in structural brain changes, and this has been reported in many conditions.[17, 18] In cases where the central visual field defects in MD overlap in the two eyes, this results in a reduced stimulation of part of the visual pathways and visual cortex.[19, 20] This, in turn, could result in changes to the structure of the visual cortex.[21] Indeed, various retinal and optic nerve conditions have been found to be associated with cortical thinning or cortical grey matter loss[21-24], and with volumetric reduction of the more proximal visual pathway structures such as the optic nerves.[25] This raises the question whether MD might also affect the structural integrity of the visual pathways. This question is relevant, as a positive answer might implicate a need for modifications of the clinical management of MD. The question is also relevant in the light of the conflicting reports on the presence of functional occipital reorganization following retinal lesions.[20, 26-30] Here, we investigate whether MD affects the structural integrity of the visual pathways by comparing their structural MRI morphometry in patients with MD and healthy controls.

# METHODS

**Subjects.** Participants of this study were recruited in Groningen (the Netherlands), York and London (United Kingdom) and Regensburg (Germany). This study conformed to the tenets of the Declaration of Helsinki and was approved by the respective medical review board of each centre participating in the study, and national regulatory ethics bodies, where necessary. The inclusion criteria required that subjects must be free from neurological or psychiatric disorders, and in the case of the patients present with central visual field defect attributed to bilateral macular degeneration. All participants gave their written informed consent before participation. The participants with macular degeneration were classified into two groups: patients with JMD and patients with AMD. In total, 114 subjects participated in this study: 25 with AMD, 34 with JMD, and the remaining 55 were healthy controls. However, one patient with AMD decided to quit the study, leaving in total 113 subjects included in the analyses.

The control subjects had a mean age of 49.6 years (range 13 to 83 years), and 46% of them were males. Of these controls, 22 subjects were age-matched (mean age 68, range 61-83 years) to the AMD group. The other 33 subjects (mean age 37.4, range 13 - 60 years) were age-matched to the JMD group. The characteristics of the affected participants are described in Table 1.

| Characteristics | | Values | |
|---|---|---|---|
| | | AMD | JMD |
| Number of subjects | | 24 | 34 |
| Age, mean (range), years | | 75.2 (52 - 91) | 40.2 (12 - 66) |
| Female, sex, % | | 42 | 38 |
| Visual acuity in logMAR, mean (range) | | | |
| | OD | .96 (1.60 - .50) | 1.12 (3.00 - .30) |
| | OS | .88 (1.80 - .10) | 1.16 (3.00 - .66) |
| Scotoma diameter, mean (range), degree | | 14 (4 - 25) | 20 (3 - 65) |

**Table 1. Baseline patient characteristics.** Characteristics of the two patient groups (AMD and JMD) were age, sex proportion (in percentage), visual acuity for one or both eyes (expressed in logMAR), and scotoma diameter (expressed in degree). AMD – age-related macular degeneration; JMD – juvenile type macular degeneration; logMAR – logarithm of minimum angle of resolution; OD - right eye (oculus dexter); OS - left eye (oculus sinister).

In Table 2, the number of participants scanned at each location, and mean age of the separate groups are shown. In the remainder of the paper, we therefore consider four groups: 1) AMD patients, 2) JMD patients, 3) the group of healthy older control subjects (HCO), age-matched to the AMD patients, and 4) the group of healthy young controls (HCY), age-matched to the JMD patients.

|  | | AMD | HCO | JMD | HCY | Total |
|---|---|---|---|---|---|---|
| **Groningen** | | | | | | |
| | number of subjects | 8 | 7 | 0 | 3 | **18** |
| | Age, mean (range), years | 73.3 | 67.6 | | 56.7 | |
| | | (52 – 83) | (61 – 83) | | (56 – 58) | |
| **York** | | | | | | |
| | number of subjects | 8 | 5 | 8 | 7 | **28** |
| | Age, mean (range), years | 81.1 | 68.4 | 33.1 | 27.1 | |
| | | (71 – 91) | (61 – 78) | (20 – 50) | (19 – 38) | |
| **Regensburg** | | | | | | |
| | number of subjects | 8 | 10 | 26 | 23 | **67** |
| | Age, mean (range), years | 71.3 | 68.1 | 42.4 | 38.0 | |
| | | (54 – 83) | (61 – 83) | (12 – 66) | (13 – 60) | |
| | Disease duration, mean, years | 7.6 | | 15.9 | | |
| | | (1 – 21) | | (2 – 42) | | |
| **Total number of subjects** | | 24 | 22 | 34 | 33 | 113 |

**Table 2. Distribution of the four subject groups to the three different scanner locations.** The table shows the number of subjects, mean age and, if available, mean duration of the disease in the specific subject group scanned at each location. Abbreviations: JMD - juvenile macular degeneration; HCY - healthy controls young (age-matched to JMD); AMD - age-related macular degeneration; HCO - healthy controls old (age-matched to AMD).

**Magnetic resonance image acquisition.** Since the datasets were from three different study centres, the magnetic resonance (MR)-acquisition parameters varied to some extent. However, all of the acquisitions were of 1 mm x 1 mm x 1 mm resolution. MR imaging was performed on three different scanners. Groningen datasets were acquired on a 3.0 Tesla Philips Intera (Eindhoven, The Netherlands) at the MRI Centrum, University Medical Center Groningen. A three-dimensional structural image was acquired on each subject using a sequence T1W/3D/TFE-2, 8° flip angle, repetition time 8.70 ms, matrix size 256 x 256, field of view 230 x 160 x 180, yielding 160 slices. The York dataset was acquired using 8-channel, phase- array head coils on either a Siemens Trio 3 Tesla at the Combined Universities Brain Imaging Center (Royal Holloway University of London). Multi-average, whole-head T1-weighted anatomical volumes were acquired for each participant. Sequences used were MDEFT 16° flip angle, repetition time 7.90 ms, matrix size 256 x 256, field of view 176 x 256 x 256, yielding 176 sagittal slices. The Regensburg dataset was acquired on a 3.0 Tesla Allegra Scanner (Siemens, Erlangen, Germany). One hundred and sixty slices covering the whole brain, field of view = 256 x 256 mm, were obtained from each subject, using the Alzheimer's Disease Neuroimaging Initiative (ADNI)[31] sequence (TR = 8.79 ms, TE = 2.6 ms, flip angle 9°).

**Data preprocessing.** *Bias correction and noise reduction.* The images were converted from their native formats into analyse (NIFTI) format. We performed bias correction (implemented in the SPM8 segmentation tool) and noise reduction using SUSAN (Smallest Univalue Segment Assimilating Nucleus) (Smith and Brady, 1997) prior to the next steps. SUSAN is a part of FSL (FMRIB Software Library, http://www.fmrib.ox.ac.uk/fsl).

*Rigid body registration.* We performed rigid body registration on the brains to a common template using SPM8's tool for co-registration. In this step, the brains are reoriented into the common template space using six linear transformation parameters: 3 translations and 3 rotations. All transformations are performed within a 3-dimensional coordinate system, with x, y, and z as its axes. This rendered the images to have uniform dimensions and to be in approximate alignment to each other.

*Segmentation, registration, and modulation.* We used SPM8's DARTEL (Diffeomorphic Anatomical Registration through Exponentiated Lie Algebra) suite of tools.[32, 33] DARTEL tools enabled us to create modulated grey and white matter images that were registered to a common reference image specifically representing our sample. The study-specific approach we used here enabled a more accurate inter-subject registration of brain images with improved localization and sensitivity of the voxel-based morphometry (VBM). The process began with SPM8's segmentation, which segmented the co-registered $T_1$-weighted images, except for lateral geniculate body extraction. For the latter structure, we used FAST (FMRIB's Automated Segmentation Tool) and fed the output into the DARTEL pipeline. After all the brains were segmented, a reference –or template– image was generated. The first step in generating this reference image was averaging the images of all brains. Following this, the individual brains were deformed and registered as closely as possible to this reference image. Next, using the registered brain images, a new average reference image was created to which the individual brain images were again registered. After 6 of these averaging and registration cycles, the final reference image was generated. The final reference image was then used as the template to which the native segmentations of the individual brains in the study were registered and modulated.

**Data analysis.** We performed two types of analyses: (1) a whole brain voxel-wise analysis; and (2) a region-of-interest (ROI)-based analysis. In the first type of analysis we did a whole brain analysis to examine whether there were volumetric differences in general in AMD and JMD patients. In the second type of analysis, we used masks to single out the visual pathways from the brain and analyze data specifically in those regions.

*Voxel-wise analysis.* The process from the segmentation to the voxel-wise statistical analyses is known as VBM, and is implemented in the SPM8 software package (Wellcome Department of Imaging Neuroscience, London, UK; http://www.fil.ion.ucl.ac.uk/spm).[34] VBM statistically assesses local changes in grey and/or white matter volume between groups of anatomical scans, and makes it possible to spatially detect any deviation in the visual pathways. Comparisons using ANCOVA were done separately for the AMD group to their specific group of age-matched controls, and for the JMD to their specific group of age-matched controls. Age was entered as a covariate in the analysis. Threshold-free cluster enhancement (TFCE) methods[35] were applied to minimize the need for large scale smoothing or for predefining a significant cluster size. To control for a scanning site effect, we included scanner location as a covariate in the voxel-wise analyses. Differences in volumes between patients and healthy control subjects are shown in the figures for voxels with an associated $p<.01$ for the whole brain analysis, and $p<.05$ for the masked voxel-wise analysis (uncorrected for multiple comparisons). Only clusters with a minimum of 10 voxels are displayed.

**ROI definition.** In the ROI-based analysis, we used masks to extract ROI volumes from the modulated normalized brain segments. We used masks for the pregeniculate structures, the lateral geniculate bodies, the geniculo-calcarine radiations, the occipital pole, and the intracalcarine and supracalcarine cortices. The masks that we used to extract the volume of the pregeniculate structures and the lateral geniculate bodies were created manually, as the boundaries of those structures can be clearly defined visually. These visually defined masks were created on the average of all the images. The mask for the geniculo-calcarine radiations was taken from the Jülich histological atlas[36, 37], whereas the mask that covered the occipital pole, intra- and supracalcarine cortices were taken from the Harvard Center for Morphometric Analysis.[38-42] We defined five regions-of-interest (ROIs) along the visual pathway: pregeniculate structures (PGCL), which include the optic nerve, the chiasm and the optic tract, lateral geniculate bodies (LGB), geniculocalcarine radiations (GCR), also known as optic radiations, the occipital pole (OCP) and the calcarine region (CCR), which contains the intracalcarine and supracalcarine cortices. These ROIs mark the separate volumes along the visual pathway we were interested in. Figure 1 shows the locations of the masks used to define the ROIs on a template brain.

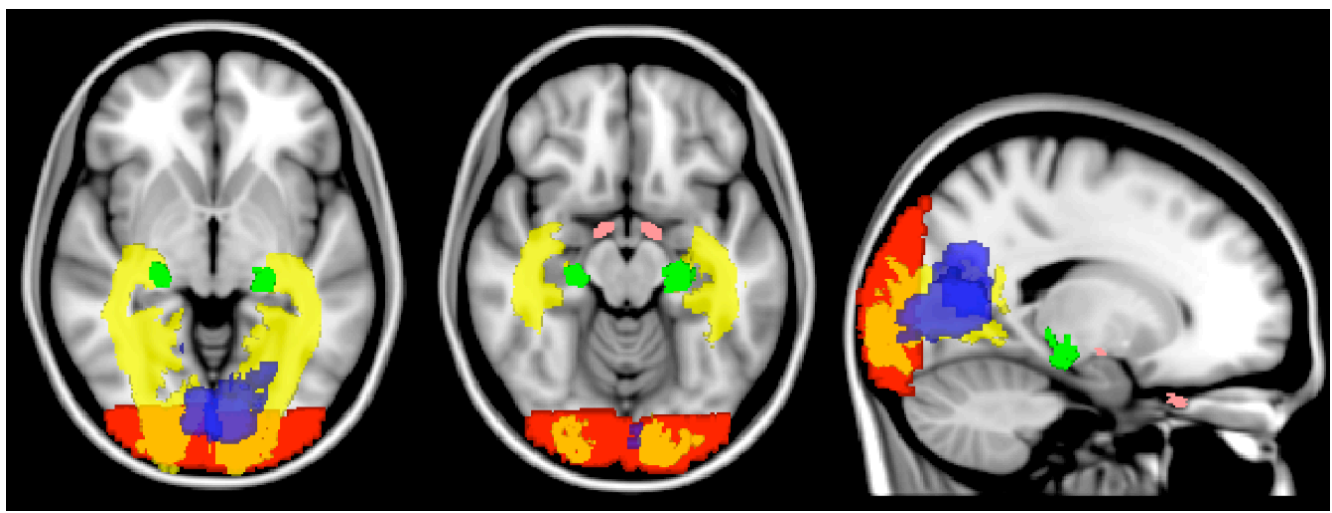

**Figure 1. Location of the regions of interest located along the visual pathways.** Colour codes for the different regions of interest: Pink – Pregeniculate structures (PGCL); Green – Lateral geniculate bodies (LGB); Yellow – Geniculocalcarine radiations (GCR); Red – occipital pole (OCP); Blue – Calcarine region (CCR).

Statistical analysis of the ROI data. Prior to the statistical analysis, we verified that the extracted values of the ROIs were normally distributed with the Kolmogorov-Smirnov test. Using repeated measures ANOVA, we analyzed the significance of the differences in volume of the entire visual pathway between the patient groups and their respective age-matched control groups. In this analysis, group was entered as a between-subject variable and ROI as a within-subject variable. Age and scanner location were added as a covariate. The significance of the differences in volume of the separate ROIs were analyzed with t-tests. We visualized these differences within the defined ROIs using a masked voxel-wise analysis.

## RESULTS

**Whole brain voxel-wise analysis.** We performed a whole brain voxel-wise analyses to examine whether there were differences in the grey and the white matter in MD in the visual pathways, but also in regions outside of the visual pathways. Figure 2 shows results for this analysis. The upper row of images shows the results for the JMD group, while the lower row of images is for the AMD group, compared to their respective age-matched healthy controls. In both the JMD and the AMD groups, we found reductions of grey (green and yellow color coded, respectively) and white (red and blue color coded, respectively) matter located in the visual cortex and the optic radiations. The reductions, in particular in the white matter, appear more pronounced in the AMD group than for the JMD group. In the AMD group, we also find marked differences outside of the visual pathways. These were mainly located in white matter in the frontal lobe, as shown in Figure 2, compared to the age-matched healthy controls ($p<.01$ uncorrected).

We did not find any increases in grey or white matter in the whole brain voxel-wise analyses for both types of MD.

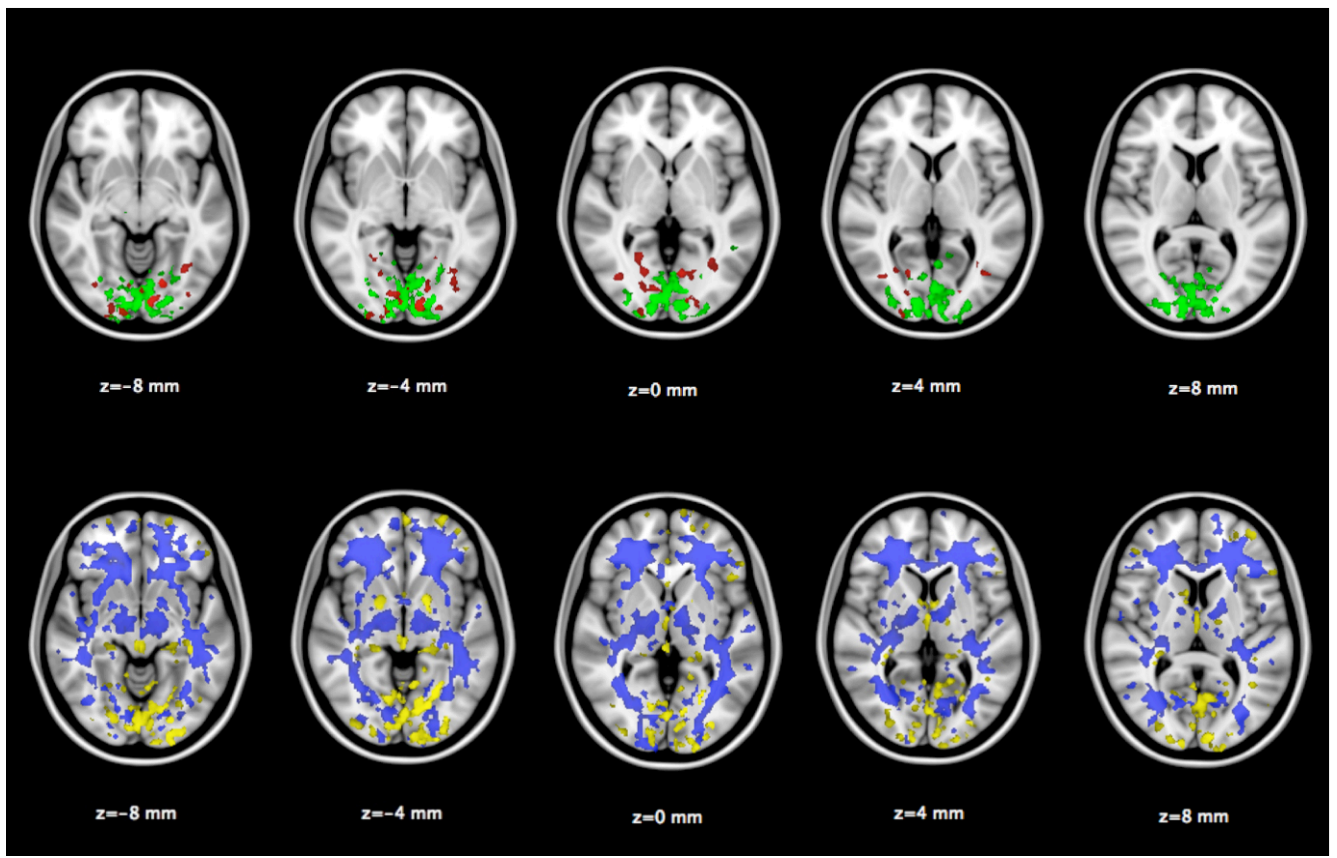

**Figure 2. Volumetric reductions in juvenile (JMD) and age-related (AMD) macular degeneration.** The transversal slices on the upper row highlight the areas where the JMD group shows significantly lower volume compared to the age-matched controls ($p<.01$, uncorrected). Green shows the involvement of grey matter, and red shows the involvement of white matter. The transversal slices on the lower row highlight the areas where the AMD group shows significantly lower volume compared to the age-matched controls ($p<.01$, uncorrected). Yellow shows the involvement of grey matter, blue shows the involvement of white matter. Talairach position of the slices is given by their "z" value.

**ROI-based analysis.** In both the patients and the controls groups, the distribution of the volumes of the ROIs did not deviate from normality. Table 3 shows, for the five ROIs, the mean volume and the relative differences of the ROI volume for the patient groups relative to their age-matched controls. Next, we tested whether there were volumetric differences in the defined ROIs along the visual pathway between patients and controls. Figure 3 shows the mean volumes of each ROI for all four groups: the AMD patients, the JMD patients, the older healthy controls and the younger healthy controls. Figure 4 shows the distribution of the ROI volume as a function of age for the various ROIs. The regression lines show the effect of age for the two control groups combined. In PGCL and GCR, there is no significant correlation between ROI volume and age in the healthy controls.

In the other ROIs – LGB, OCP and CCR – volume reduces with age. The Pearson correlation coefficients are respectively -.310, -.223 and -.252, in the LGB and CCR these correlation are significant (p<.05). We therefore included age as a covariate in our subsequent analyses.

|  | ROI $\mu \pm \sigma_\mu$ | | | | |
| --- | --- | --- | --- | --- | --- |
|  | PGCL (mm³) | LGB (mm³) | GCR (cm³) | OCP (cm³) | CCR (cm³) |
| AMD | 401.58 ± 21.94 | 86.67 ± 3.17 | 9.51 ± .23 | 5.72 ± .12 | 6.66 ± .20 |
| HCO | 459.41 ± 24.75 | 91.77 ± 3.02 | 11.29 ± .17 | 6.64 ± .12 | 8.09 ± .29 |
| JMD | 382.62 ± 13.44 | 95.94 ± 3.48 | 10.78 ± .18 | 6.35 ± .11 | 7.91 ± .16 |
| HCY | 419.06 ± 12.64 | 109.76 ± 3.49 | 11.68 ± .19 | 6.95 ± .15 | 8.58 ± .23 |
|  | Relative difference to the controls (%) | | | | |
| AMD | -12.6 | -5.6 | -15.8 | -13.9 | -17.7 |
| JMD | -8.7 | -12.6 | -7.7 | -8.6 | -7.8 |

**Table 3. ROI morphometry.** Morphometric values in term of area or volume were extracted from ROIs situated along the visual pathway. ROI - regions of interest; $\mu$ - mean; $\sigma\mu$ - standard errors of mean; AMD - age-related macular degeneration; HCO - healthy controls old (age-matched to AMD); JMD - juvenile macular degeneration; HCY - healthy controls young (age-matched to JMD); PGCL - pregeniculate structures; LGB - lateral geniculate bodies; GCR - geniculocalcarine radiation; OCP - occipital poles; CCR – calcarine region.

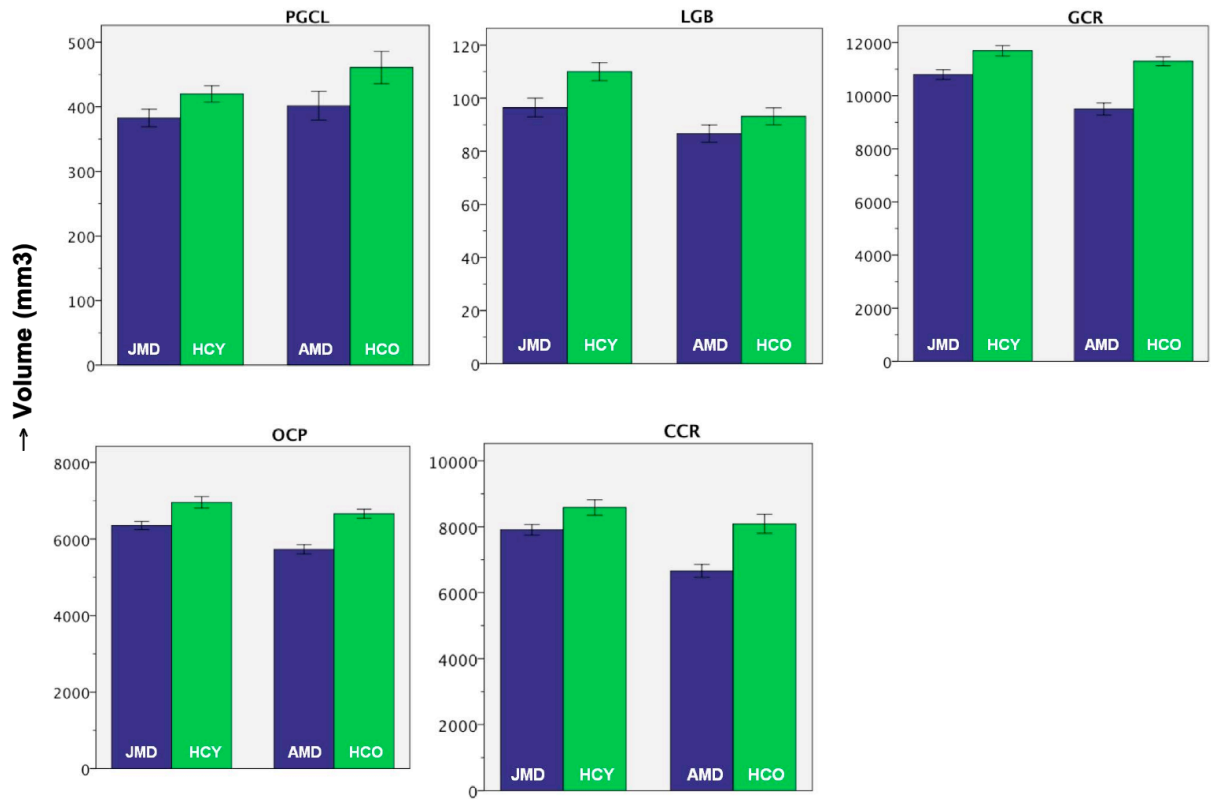

**Figure 3. Mean volume of five ROIs located along the visual pathways.** ROI – region of interest; JMD – patients with juvenile macular degeneration; HCY - healthy controls young (age-matched to JMD); AMD – patients with age-related macular degeneration; HCO – healthy controls old (age-matched to AMD);

ROI abbreviations: PGCL – pregeniculate structures; LGB – lateral geniculate bodies; GCR - geniculocalcarine radiation; OCP - occipital poles; CCR – calcarine region. The bars show the mean value of the ROI-volume in the particular group. The error bars show +/- 1 standard error of the mean.

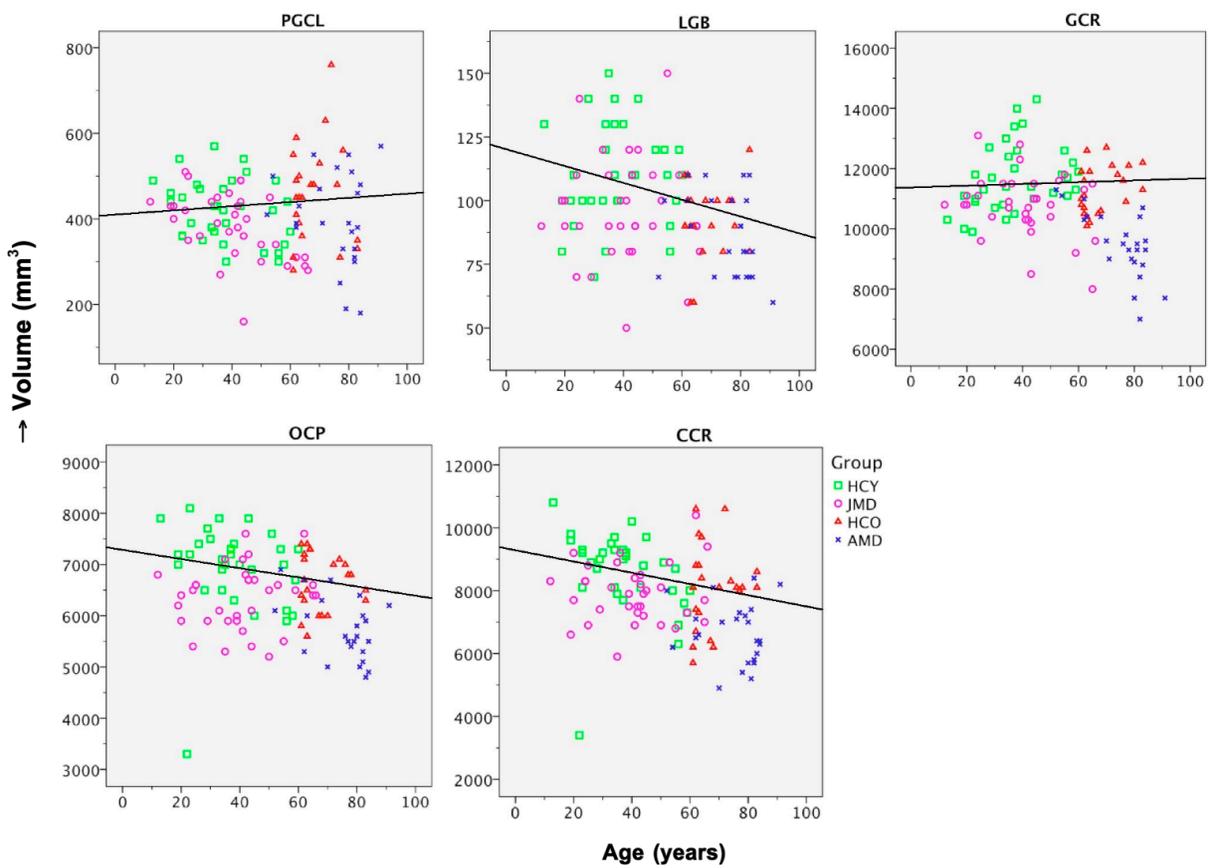

**Figure 4. ROI volume as a function of age.** ROI – region of interest; HCY - healthy controls young (age-matched to the JMD-patients); JMD – patients with juvenile macular degeneration; HCO - healthy controls old (age-matched to the AMD-patients); AMD - patients with age-related macular degeneration. ROI abbreviations: PGCL – pregeniculate structures; LGB – lateral geniculate bodies; GCR - geniculocalcarine radiation; OCP - occipital poles; CCR – calcarine region.

In general, for all ROIs, the volume is lower in the AMD patients than in the age-matched control group ($F(1, 42) = 20.415$, $p<.001$). The same is true for the volume of the ROIs in the JMD compared to the age-matched control group ($F(1, 63) = 26.170$, $p<.001$). T-tests showed that for the AMD vs. HCO groups the differences were significant in the GCR, OCP and CCR ROIs, and for the JMD vs. HCY the differences were significant in the LGB, GCR and OCP ROIs (all $p<0.01$). Table 4 shows all statistical values for the comparisons of ROI volumes between patients and controls.

| Comparison | | Regions of Interest | | | | |
|---|---|---|---|---|---|---|
| | | PGCL | LGB | GCR | OCP | CCR |
| **AMD vs. HCO** | *t*-value (*df* = 44) | 1.783 | 1.421 | 6.310 | 5.458 | 4.153 |
| | *p*-value | .082 | .162 | <.001 | <.001 | <.001 |
| **JMD vs. HCY** | *t*-value (*df* = 65) | 2.018 | 2.759 | 3.381 | 3.301 | 2.408 |
| | *p*-value | .048 | .008 | .001 | .002 | .019 |

**Table 4. Statistics for comparison of ROI volumes between patients and controls.** The upper part shows the statistical values for the comparison of the AMD group to the HCO group, whereas the lower part shows the statistical values for the comparison of the JMD group to the HCY group (t-values, degrees of freedom and p-values are given). Abbreviations: JMD - juvenile macular degeneration; HCY - healthy controls young (age-matched to JMD); AMD - age-related macular degeneration; HCO - healthy controls old (age-matched to AMD); PGCL - pregeniculate structures; LGB - lateral geniculate bodies; GCR - geniculocalcarine radiation; OCP - occipital poles; CCR – calcarine region.

In the subgroup of patients from Regensburg, data on duration of the macular degeneration at the time of scanning, was available. This subgroup from Regensburg contained 34 patients in total, of which 26 were JMD patients and 8 were AMD patients. In this group, we did not find a significant correlation between ROI volume and disease duration in AMD or in JMD.

**ROI-masked voxel-wise visualization.** To visualize the distribution of the volumetric differences in the ROIs, voxel-wise statistical comparisons were done on the ROI brain regions of the visual pathway of AMD and JMD patients. **Figure 5. Volumetric reductions in the masked visual pathway ROIs in age-related macular degeneration (AMD).** The transversal slices of the template brain highlight the areas where the AMD group shows significantly lower volume compared to the age-matched controls (p<.05, uncorrected), within the defined ROI masks along the visual pathway. Involvement of the grey matter (yellow) can be seen in the visual cortex, whereas white matter involvement (blue) can be seen in the optic tract and in the regions corresponding to the geniculocalcarine radiation. The Talairach position of the slices is given by the "z" value.

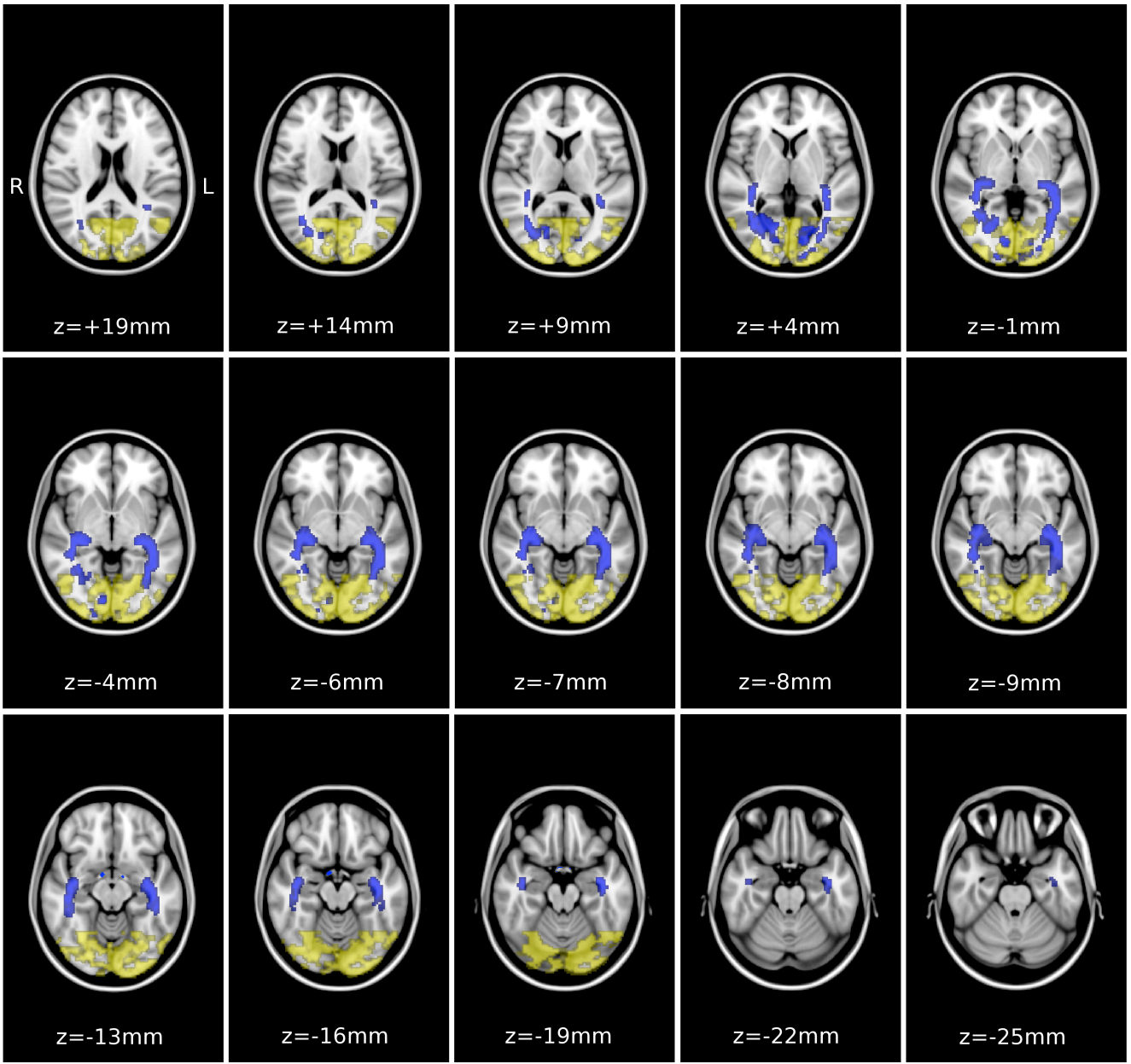

Figure 5 shows a comparison between AMD patients and age-matched controls, whereas Figure 6 shows the comparison for the JMD patients and age-matched controls. For this visualization, a rather liberal threshold of p<.05 (uncorrected) was used, to indicate the extent of the changes within the ROIs.

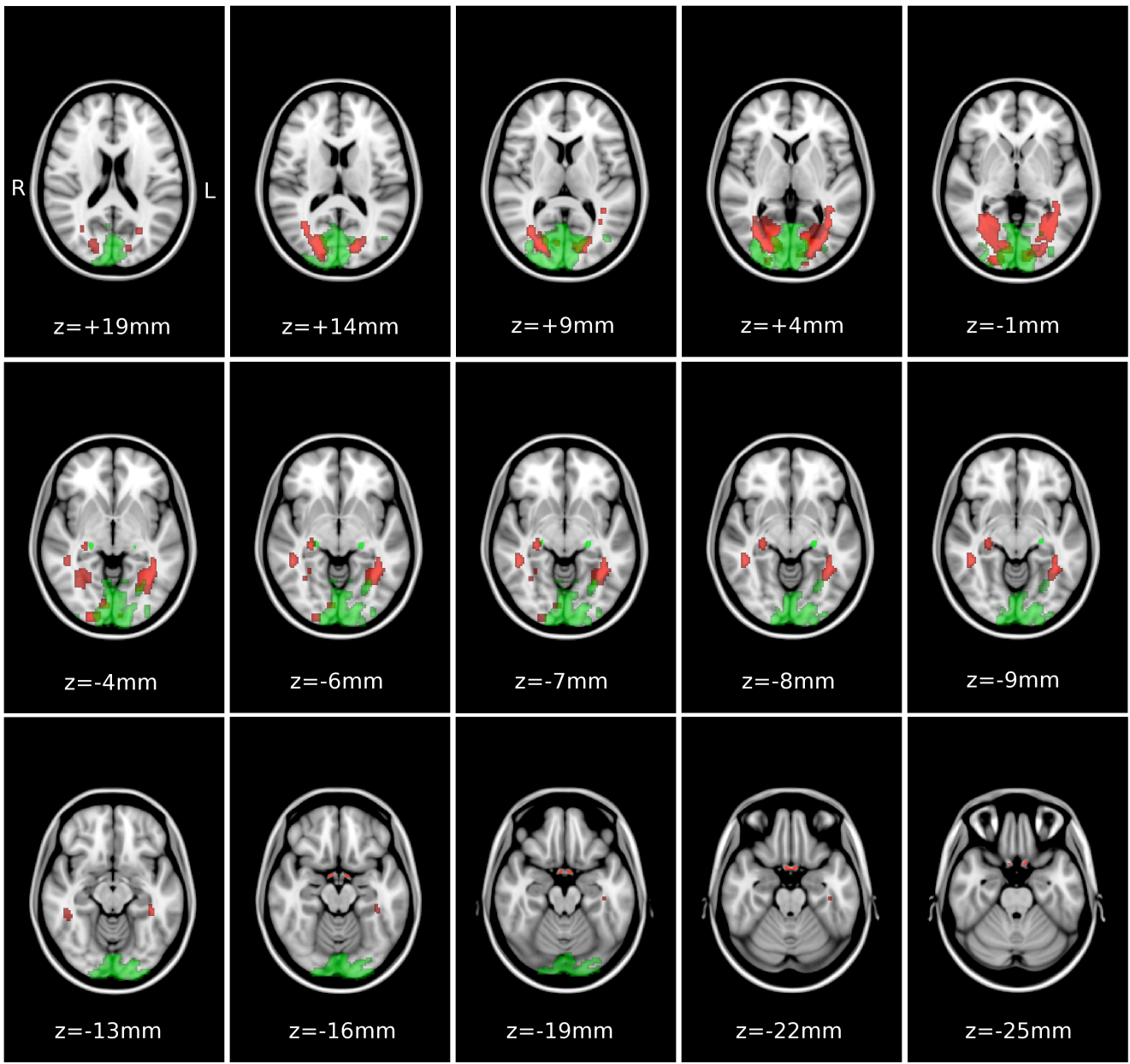

**Figure 6. Volumetric reductions in the masked visual pathway ROIs in juvenile macular degeneration (JMD).** The transversal slices of the template brain highlight the areas where the JMD group shows significantly lower volume compared to the age-matched controls (p<.05, uncorrected), within the defined ROI masks along the visual pathway. Involvement of the grey matter (green) can be seen in the visual cortex and at the position of the lateral geniculate bodies, whereas white matter involvement (red) can be seen in the pregeniculate structures and in regions corresponding to the geniculo-calcarine radiation. The Talairach position of the slices is given by the value "z".

In both types of MD, the volumetric reductions identified by our ROI analysis are dispersed throughout all of the retrobulbar afferent visual pathway structures.

# DISCUSSION

MD is known as one of the primary causes of blindness and its retinal pathology has been studied extensively. Only in recent years, however, the involvement of the cerebral cortex in this pathology has become a topic of study.[20, 21, 26, 43-45] Our previous findings of grey matter loss in the visual cortex in AMD and glaucoma[21], as well as in JMD[24], and a reduced volume of the visual pathways in glaucoma[25] prompted us to study the question whether the (pre-cortical) visual pathway in MD might also be affected. Here, we found that in MD there was indeed volumetric loss of the visual pathway structures. Moreover, in the whole brain voxel-wise analysis we found a volumetric reduction in the frontal lobe in AMD patients, mainly in the white matter. While previous studies have shown that in MD the brain is functionally and structurally affected[19, 21, 24], to our knowledge, this is the first one to report on the extensiveness of the structural changes.

What may have caused this volumetric loss of the pathway structures in MD? A similar loss has previously been reported in glaucoma.[25] In glaucoma, it is rather intuitive that the retrobulbar visual pathway may be involved in its pathology as there is a loss of retinal ganglion cells (RGCs) and their axons form the optic nerve. However, in MD, the primary loss of vision is due to damage of the photoreceptors[46], which are first-order neurons. This degeneration does not necessarily have to be transferred across the retinal layers to the bipolar (second-order neuron) and RGC (third-order neuron) layers and their projecting axons. This argument would hold even more for the geniculate and post-geniculate structures. However, studies indicate that a loss of RGCs may also be associated with the loss of photoreceptors in both wet AMD[47] as well as in the later stages of dry AMD (geographic atrophy).[43, 48] This indicates that RGCs can be affected by the adverse effect of photoreceptor degeneration and provides a possible mechanism that can explain the volumetric reduction of the visual pathway and cortex in patients with MD.

Our initial ROI-based analysis, in which we investigate the volume of the separately defined structures of the visual pathways, indicated changes in both AMD and JMD, when compared to the age-matched controls (Figure 3, and Table 3). We also plotted ROI volume against age to verify the presence of a correlation with age. Indeed, in the healthy controls we found a negative correlation of age and volume to be present in the LGB, OCP and CCR (Figure 4). To account for such age effects, age was included as a co-variate in the subsequent ROI- and voxel-wise analyses. In the voxel-wise analyses, we compared the brains voxel-by-voxel along the visual pathway. In JMD, we found volumetric differences along the full extent of the visual pathways. In AMD we primarily found volumetric differences in the optic radiations and the visual cortex. Moreover, we could confirm the previous findings of a reduction in grey matter in the early visual cortex.

Differences in PGCL and LGB in AMD were also shown in the uncorrected masked voxel-wise analysis, but these findings did not remain significant in the ROI-analysis. This difference in results between JMD and AMD group can have various possible explanations. It could be caused by the lower number of AMD patients, compared to JMD patients, respectively 24 and 34. Furthermore, it could be associated with notion that the volume of the PGCL appears to be larger in the older control group than in the younger control group, as shown in Table 3. This apparent increase in pre-geniculate volume with age might be a factor in the difference in results between AMD and JMD, although this difference in PGCL volume between HCO and HCY is not significant.

In both AMD and JMD the volume of the optic radiations are reduced markedly. The effect in optic radiations actually appears to be larger than the effect in the anterior visual pathways. A possible reason for this could be that besides feedforward projections from the LGB to the visual cortex, corticogeniculate neurons provide input from the visual cortex back to the LGB.[49] Such feedback pathways have not been found between LGB and retina. Therefore, the optic radiations could be affected more than the anterior visual pathways. Besides that, the size of the GCR ROI is much larger than the PGCL, which implies that there might be more overlap between the subjects in the GCR ROI than in the PGCL ROI.

Even though the visual cortex is one of the regions of cerebral cortex that is least affected by age[50] the age effects in the control groups suggest that it is necessary to account for age in the analyses. Duration of the disease could possibly also have an effect on the degree of changes in the visual pathways. In this study there was data available on disease duration only for the participants from Regensburg. In this subgroup, we did not find a significant correlation between disease duration and ROI volume. Nevertheless, we suggest that future studies should still consider taking the duration of disease into account. JMD starts earlier in life and thus, at a particular age of the subject, the disease period is usually longer, and thus hypothetically JMD will have had a larger chance than AMD to affect the retrobulbar visual pathway.

Unexpectedly, in the whole brain voxel-wise analysis, we also found a lower volume in white matter in the frontal lobe, specifically in the AMD participants. As far as we know, no relationship between AMD and frontal lobe volume has been described. This finding could be associated with a possible link between AMD and mild cognitive impairment and Alzheimer's disease, which has been suggested earlier, based on a number of epidemiological studies.[51, 52] It should be noted, however, that smoking is an important co-factor in this correlation.[51] Unfortunately, we did not have information about smoking habits on our patients. Nevertheless, a common pathogenic mechanism for AMD and Alzheimer's disease has been described.[53] It is less plausible that reduced frontal lobe volume could be due to a decreased activity of AMD participants as a consequence of their visual impairment. If so, we would expect to find a similar effect in the JMD participants, which was not the case. We suggest that a follow-up study is required to confirm the presence of an association between AMD and frontal lobe grey and white matter density.

Our study design was cross-sectional, therefore we can[54] not determine which happened first, the volumetric reduction in the brain, or the degeneration of the macula. Macular degeneration has been managed clinically as an ocular pathology. However, the structural changes, mainly in the lateral geniculate bodies and the visual cortex, may alert clinicians to the possibility that brain-related mechanism are involved in the disease. A neuroprotective approach to preserve the photoreceptors and retinal pigment epithelium by the administration of ciliary neurotrophic factors[55] or anti-inflammatory[56] drugs might be worthwhile to consider as a first step. Our findings suggest that any retinal neuroprothesis[57] should be implanted as early as possible to avoid long-term, and perhaps irreversible, reductions in grey matter of the visual pathway. A better comprehension of the pathology underlying the various macular degeneration types and diagnostic approach are of utmost importance to allow the quantification of treatment outcome. We believe that neuroimaging and retinal imaging[58], as well as their quantification methods[21, 34, 59, 60], could – over time – become clinically viable options to evaluate the progress of the disease and the treatment outcome.

In summary, in both age-related and juvenile MD the volume of the visual pathway is significantly reduced. The extent of involvement was different for the two types of disease, with JMD showing the most widespread involvement of the visual pathway. This implies that it could be useful to also investigate the chronology of the visual pathway changes in macular degeneration. Moreover, we found an indication that the white matter volume in the frontal lobe is reduced in AMD. This could be the neural correlate of the proposed link between AMD and mild cognitive impairment and Alzheimer's disease.